\documentclass{wscpaperproc}
\usepackage{latexsym}
\usepackage{graphicx}
\usepackage{mathptmx}
\usepackage[T1]{fontenc}
\usepackage{amsmath}
\usepackage{amsthm}
\usepackage{amsfonts}
\usepackage{amssymb}
\usepackage{booktabs}
\usepackage{tikz}
\usetikzlibrary{arrows.meta,positioning,fit,backgrounds,calc,decorations.pathreplacing}
\usepackage[pdftex,colorlinks=true,urlcolor=blue,citecolor=black,anchorcolor=black,linkcolor=black]{hyperref}
\usepackage[linesnumbered,ruled,vlined]{algorithm2e}

\SetKwInOut{Input}{Input}
\SetKwInOut{Output}{Output}
\SetKwComment{Comment}{$\triangleright$\ }{}
\DontPrintSemicolon

\newtheoremstyle{wsc}{3pt}{3pt}{}{}{\bfseries}{}{.5em}{}
\theoremstyle{wsc}

\newcommand{\calL}{\mathcal{L}}
\newcommand{\calF}{\mathcal{F}}
\newcommand{\EE}{\mathbb{E}}

\begin{document}

% setting up general page style
\pagestyle{fancyplain}

% setting up page style of first page
\thispagestyle{plain}
\firstPageHead{}

% setting up running header (authors) of subsequent pages
\chead{\fancyplain{}{\itshape Li, Li, Chen, and Chew}}

% setting up seperation parameters
%\headsep=72pt
\rhead{}
\cfoot{}
\renewcommand{\headrulewidth}{0pt} % (renewcommand needed in fancyhdr to remove top decorative line)
%\headrulewidth=0pt  % ("setlength" needed in fancyheading to remove top decorative line)

%%%%%%%%%%%%%%%%%%%%%%%%%%%%%%%%%%%%%%%%%%%%%%%%%%%%%%%%%%%%%%%%%%%%%%%%%%%%%%
%                                                                            %
%     THESE COMMANDS ARE REQUIRED TO WORK WITH WSC.BST TO MAKE BIBLIO     %
%                                                                            %
%%%%%%%%%%%%%%%%%%%%%%%%%%%%%%%%%%%%%%%%%%%%%%%%%%%%%%%%%%%%%%%%%%%%%%%%%%%%%%
\makeatletter
\let\@internalcite\cite
\def\cite{\def\@citeseppen{-1000}%
    \def\@cite##1##2{(##1\if@tempswa , ##2\fi)}%
    \def\citeauthoryear##1##2##3{##1 ##3}\@internalcite}
\def\citeNP{\def\@citeseppen{-1000}%
    \def\@cite##1##2{##1\if@tempswa , ##2\fi}%
    \def\citeauthoryear##1##2##3{##1 ##3}\@internalcite}
\def\citeN{\def\@citeseppen{-1000}%
%  Pierre L'Ecuyer's fix for multiple cite bug
%  Added by Paul J Sanchez on 4 October 2001
%   \def\@cite##1##2{##1\if@tempswa , ##2)\else{)}\fi}%
%   \def\citeauthoryear##1##2##3{##1 (##3}\@citedata}
    \def\@cite##1##2{##1\if@tempswa, ##2)\else{}\fi}%
    \def\citeauthoryear##1##2##3{##1 (##3)}\@citedata}
\def\citeA{\def\@citeseppen{-1000}%
    \def\@cite##1##2{(##1\if@tempswa , ##2\fi)}%
    \def\citeauthoryear##1##2##3{##1}\@internalcite}
\def\citeANP{\def\@citeseppen{-1000}%
    \def\@cite##1##2{##1\if@tempswa , ##2\fi}%
    \def\citeauthoryear##1##2##3{##1}\@internalcite}
\def\shortcite{\def\@citeseppen{-1000}%
    \def\@cite##1##2{(##1\if@tempswa , ##2\fi)}%
    \def\citeauthoryear##1##2##3{##2 ##3}\@internalcite}
\def\shortciteNP{\def\@citeseppen{-1000}%
    \def\@cite##1##2{##1\if@tempswa , ##2\fi}%
    \def\citeauthoryear##1##2##3{##2 ##3}\@internalcite}
\def\shortciteN{\def\@citeseppen{-1000}%
%  Pierre L'Ecuyer's fix for multiple cite bug
%  Added by Paul J Sanchez on 2 September 2002
%  should have caught this last year...
%   \def\@cite##1##2{##1\if@tempswa , ##2)\else{)}\fi}%
%   \def\citeauthoryear##1##2##3{##2 (##3}\@citedata}
% Shane G. Henderson fix for extra right bracket at end of optional material June 8, 2005
%    \def\@cite##1##2{##1\if@tempswa, ##2)\else{}\fi}%
    \def\@cite##1##2{##1\if@tempswa, ##2\else{}\fi}%
    \def\citeauthoryear##1##2##3{##2 (##3)}\@citedata}
\def\shortciteA{\def\@citeseppen{-1000}%
    \def\@cite##1##2{(##1\if@tempswa , ##2\fi)}%
    \def\citeauthoryear##1##2##3{##2}\@internalcite}
\def\shortciteANP{\def\@citeseppen{-1000}%
    \def\@cite##1##2{##1\if@tempswa , ##2\fi}%
    \def\citeauthoryear##1##2##3{##2}\@internalcite}
\def\citeyear{\def\@citeseppen{-1000}%
    \def\@cite##1##2{(##1\if@tempswa , ##2\fi)}%
    \def\citeauthoryear##1##2##3{##3}\@citedata}
\def\citeyearNP{\def\@citeseppen{-1000}%
    \def\@cite##1##2{##1\if@tempswa , ##2\fi}%
    \def\citeauthoryear##1##2##3{##3}\@citedata}
%
% \@citedata and \@citedatax:
%
% Place commas in-between citations in the same \citeyear, \citeyearNP,
% \citeN, or \shortciteN command.
% Use something like \citeN{ref1,ref2,ref3} and \citeN{ref4} for a list.
%
\def\@citedata{%
    \@ifnextchar [{\@tempswatrue\@citedatax}%
                  {\@tempswafalse\@citedatax[]}%
}

\def\@citedatax[#1]#2{%
\if@filesw\immediate\write\@auxout{\string\citation{#2}}\fi%
  \def\@citea{}\@cite{\@for\@citeb:=#2\do%
    {\@citea\def\@citea{, }\@ifundefined% by Young
       {b@\@citeb}{{\bf ?}%
       \@warning{Citation `\@citeb' on page \thepage \space undefined}}%
{\csname b@\@citeb\endcsname}}}{#1}}%

% don't box citations, separate with ; and a space
% also, make the penalty between citations negative: a good place to break.
%
\def\@citex[#1]#2{%
\if@filesw\immediate\write\@auxout{\string\citation{#2}}\fi%
  \def\@citea{}\@cite{\@for\@citeb:=#2\do%
    {\@citea\def\@citea{; }\@ifundefined% by Young
       {b@\@citeb}{{\bf ?}%
       \@warning{Citation `\@citeb' on page \thepage \space undefined}}%
{\csname b@\@citeb\endcsname}}}{#1}}%

% (from apalike.sty)
% No labels in the bibliography.
%
\def\@biblabel#1{}
\makeatother

%\newlength{\bibhang}
%\setlength{\bibhang}{2em}

% Indent second and subsequent lines of bibliographic entries. Taken
% from openbib.sty: \newblock is set to {}.
% \renewcommand{\refname}{REFERENCES}

\newdimen\bibindent
\bibindent=0.0em
% SEC: was \def\thebibliography#1{\section*{\refname\@mkboth
% SEC: was   {\uppercase{\refname}}{\uppercase{\refname}}}\list
\def\thebibliography#1{\section*{\refname}\list
   {}{\settowidth\labelwidth{[#1]}
   \leftmargin\parindent
   \itemindent -\parindent
   \listparindent \itemindent
   \itemsep 0pt
   \parsep 0pt}
   \def\newblock{}
   \sloppy
   \sfcode`\.=1000\relax}

           % Set up BiBTeX macros

% needed to make the tex document look more like the word counterpart :-(
\setlength{\baselineskip}{12.7pt}

\title{FROM ANNUAL THROUGHPUT TO VESSEL SCHEDULES: A STOCHASTIC GENERATOR FOR TRANSSHIPMENT HUB SIMULATION}

\author{\begin{center}
  Qiaohong Li, Haobin Li, Tianhao Chen, and Ek Peng Chew\\[11pt]
  {\small Dept. of Industrial Systems Engineering and Management, National University of Singapore, SINGAPORE}
\end{center}}

\maketitle

\vspace{-12pt}

\section*{ABSTRACT}

Simulating container transshipment hubs requires vessel arrivals reflecting cyclical liner schedules and origin-destination (OD) cargo pairing. Existing models relying on Poisson arrivals and aggregate transshipment volumes severely distort waiting-time and yard-occupancy predictions. We propose a three-phase schedule generator, calibrated entirely from public port statistics, eliminating the need for proprietary data. The framework introduces a two-level Gamma mathematical model that balances structured weekly services with operational perturbations. For cargo routing, we develop GreedyDwellFit (GDF), a fast allocation heuristic to pair transshipment batches to specific connecting services using a gravity model. Validated against Busan and Singapore mega-hubs, the model reproduces throughput and call frequencies with under 1\% error. Our results show that replacing Poisson models with this Gamma-GDF framework eliminates significant distortions in terminal performance projections, offering a robust, generalisable foundation for port simulation.

\section{INTRODUCTION}
\label{sec:intro}

Container transshipment hubs route tens of millions of twenty-foot equivalent units (TEUs) annually, and discrete-event simulation (DES) is the standard tool for stress-testing these facilities and informing capital investment decisions \shortcite{dragovic2017simulation,mpa2025stats}. The predictive fidelity of any terminal simulation, however, is bounded by the realism of its vessel traffic input model. Homogeneous Poisson processes remain the most common arrival generator despite the strongly cyclic nature of liner shipping, leading to inflated waiting-time predictions and potentially misleading capacity assessments \shortcite{vanAsperen2003arrival,rodrigues2016analysis}.

Current input models are deficient in two respects.  On the arrival
side, each liner service visits a port roughly once per week, yet the
standard Poisson generator
\shortcite{legato2001berth,yun1999simulation,canonaco2008queuing} discards
this schedule structure entirely.  On the cargo-routing side,
transshipment containers discharged from one service must catch a
particular outbound service; the time gap between these two calls sets the
yard dwell and the risk of misconnection.  Most simulations either ignore
transshipment or treat it as a single aggregate volume
\shortcite{petering2009development}, so they cannot represent these
dependencies.

This paper proposes a three-phase schedule generation framework built
around the concept of a
\emph{Line-Service-Direction}~(LSD)---a unique combination of shipping
service, port of call, and voyage direction that recurs weekly with a
predictable cargo profile.  The main contributions are:
\begin{enumerate}
  \item A \textbf{two-level Gamma arrival model} separating structural
        (between-service) from operational (within-service) variation.
  \item A \textbf{greedy OD transshipment allocation} (GreedyDwellFit,
        GDF) that pairs transshipment batches to specific connecting
        services with guaranteed convergence via adaptive threshold
        relaxation.
  \item An \textbf{unbiased throughput normalisation} that matches a
        target~$T$ in expectation (realised within $\sim$1\%) while
        isolating week-to-week volatility via a controllable~$CV^{TEU}$,
        enabling apples-to-apples scenario comparisons for capacity
        planning.
  \item A \textbf{general-purpose default configuration} where per-LSD
        throughput scales with vessel Gross Tonnage (GT) and the transshipment OD matrix
        uses a single-parameter gravity model---all derivable from open
        data, replaceable when proprietary data is available.
  \item A \textbf{calibration to Busan and cross-validation on Singapore},
        reproducing observed statistics to within 1\% error.
\end{enumerate}

Section~\ref{sec:related} reviews related work;
Section~\ref{sec:framework} describes the framework;
Section~\ref{sec:experiments} presents experiments;
Section~\ref{sec:conclusion} concludes.

\section{RELATED WORK}
\label{sec:related}

In practice, liner carriers run fixed weekly services, calling each
terminal on roughly the same day every week, so vessel arrivals are
intrinsically cyclic rather than memoryless.  At a transshipment hub most
boxes are relayed between two such services, so the gap between paired
calls---not an inland clearance time---sets the yard dwell.

A vessel schedule for hub simulation must specify \emph{when} vessels
arrive, \emph{how much} cargo they carry, and \emph{where} that cargo
goes.  We review the literature along these three dimensions.

\textbf{Arrival process modelling.}
Poisson arrivals remain the default in port simulation
\shortcite{legato2001berth,yun1999simulation,canonaco2008queuing}, yet
\shortciteN{vanAsperen2003arrival} showed Poisson gives the worst predictions
for liner traffic, \shortciteN{dragovic2017simulation}'s 50-year survey
concurred, and \shortciteN{rodrigues2016analysis} demonstrated that a wrong
interarrival distribution can inflate cost estimates by up to 566\%.
Schedule-aware alternatives---cyclic berth plans
\shortcite{hendriks2010robust}, deep-sea/feeder splits
\shortcite{jia2020simulation}, counting-process models
\shortcite{dicrescenzo2023vessels}, and AIS-based ETA prediction
\shortcite{yu2018ship}---all require the schedule as input; none
generates it from aggregate statistics.

\textbf{Transshipment flow modelling.}
At transshipment hubs, each discharged container waits for a specific
outbound vessel, coupling yard dwell to the schedule.
\shortciteN{petering2009development} tracked containers across calls but
treated transshipment as a lump sum without OD detail.
\shortciteN{bierwirth2015survey} flagged schedule generation as an unresolved
gap in over 100 berth allocation studies, and
\shortciteN{brouer2017vessel} argued for stochastic carrier-side models.

\textbf{Scenario generation.}
\shortciteN{hartmann2004generating} derived vessel counts and cargo volumes
from an annual throughput target---the same starting point we use---but
assumed Poisson arrivals and omitted transshipment OD flows.

To our knowledge, ours is the first to combine endogenous schedule
generation, a multi-level cyclic arrival model, and explicit OD
transshipment allocation.

\section{VESSEL SCHEDULE GENERATION FRAMEWORK}
\label{sec:framework}

\begin{figure}[htb]
\centering
\resizebox{\columnwidth}{!}{%
\begin{tikzpicture}[
  >=Stealth,
  node distance=0.35cm and 0.25cm,
  phase/.style={draw, rounded corners=3pt, minimum height=0.9cm,
                minimum width=2.8cm, align=center, font=\small},
  io/.style={draw, rounded corners=8pt, minimum height=0.55cm,
             minimum width=2.0cm, align=center, font=\scriptsize,
             fill=gray!12},
  arr/.style={->, thick, shorten >=2pt, shorten <=2pt},
  lbl/.style={font=\scriptsize, align=center, text width=2.6cm},
]

\node[io] (input) {$T,\; r_k,\; \overline{GT}_k,\; \tau,\; \alpha$};

\node[phase, right=0.7cm of input, fill=blue!8] (p1)
  {\textbf{Phase 1}\\[-1pt]{\scriptsize Normalisation}};

\node[phase, right=0.7cm of p1, fill=blue!8] (p2)
  {\textbf{Phase 2}\\[-1pt]{\scriptsize LSD Init +}\\[-2pt]{\scriptsize GDF}};

\node[phase, right=0.7cm of p2, fill=orange!10] (p3)
  {\textbf{Phase 3}\\[-1pt]{\scriptsize Port Call}\\[-2pt]{\scriptsize Generation}};

\node[io, right=0.7cm of p3] (output) {PortCall\\[-1pt]events};

\draw[arr] (input)  -- (p1);
\draw[arr] (p1)     -- (p2);
\draw[arr] (p2)     -- (p3);
\draw[arr] (p3)     -- (output);

\node[lbl, below=0.50cm of $(p1.east)!0.5!(p2.west)$]
  {$N_k,\;\EE[Q^{tl}_j]$};
\node[lbl, below=0.50cm of $(p2.east)!0.5!(p3.west)$]
  {$\calL,\;\tau_l,\;\bar{q}^x_l,$\\flow matrix $F$};

\node[font=\scriptsize\itshape, above=0.35cm of $(p1.north)!0.5!(p2.north)$]
  (offl) {offline (once)};
\draw[decorate, decoration={brace, amplitude=4pt, raise=2pt}]
  (p1.north west) -- (p2.north east);

\node[font=\scriptsize\itshape, above=0.35cm of p3.north]
  (runt) {runtime (each call)};
\draw[decorate, decoration={brace, amplitude=4pt, raise=2pt}]
  (p3.north west) -- (p3.north east);

\end{tikzpicture}%
}
\caption{Overview of the vessel schedule generation pipeline.  Phases~1--2
  execute once to build the weekly LSD template and transshipment flow
  matrix; Phase~3 executes at each simulated port call.}
\label{fig:architecture}
\end{figure}
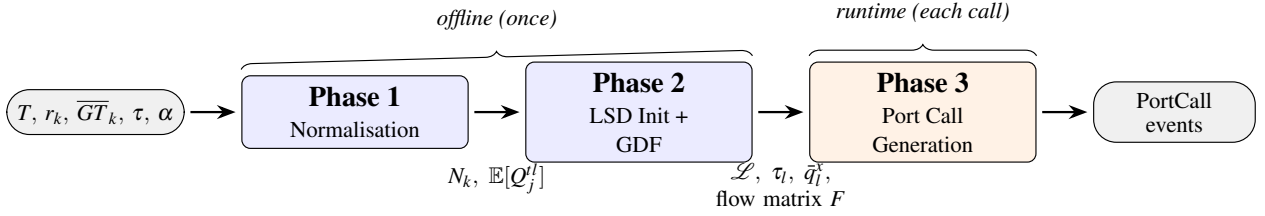

The framework builds a weekly vessel traffic schedule in three phases
(Figure~\ref{fig:architecture}).  Phase~1 converts an annual throughput
target~$T$ into per-category LSD counts.  Phase~2 gives each LSD its
attributes---arrival time, vessel size, cargo profile---and allocates
transshipment flows across LSD pairs.  Phase~3 runs during simulation: it
fires stochastic port call events, each carrying TEU volumes ready for the
terminal simulator.

\subsection{Notation and Inputs}
\label{sec:notation}

In practice, a port planner knows the annual throughput, a rough
vessel-size breakdown, and the transshipment share, but rarely a complete
vessel-by-vessel schedule.  The framework's \emph{structural parameters}
correspond to these commonly available quantities;
\emph{calibration constants} have sensible defaults that can be overridden
with better data.  Table~\ref{tab:notation} lists both.

\begin{table}[htb]
\centering
\caption{Framework inputs and distribution families.
  ``S'' = structural parameter; ``C'' = calibration constant
  with a default.  Distribution rationale: Triangular for
  physically bounded quantities; Gamma for positive-support
  tunable dispersion; Lognormal for heavy-tailed multiplicative
  variability.}
\label{tab:notation}
\footnotesize
\setlength{\tabcolsep}{3.5pt}
\begin{tabular}{lllll}
\hline
Symbol & Description & Type & Distribution \\
\hline
$T$                     & Annual throughput (TEU/yr) & S & --- \\
$K$                     & Number of vessel categories & S & --- \\
$r_k$                   & Category $k$ ratio among LSDs & S & --- \\
$\mathcal{D}^{LOA}_k$   & LOA (m) for category $k$ & S & Triangular \\
$\overline{GT}_k$       & Representative GT for category $k$ & S & --- \\
$\tau$                  & Transshipment share & S & --- \\
$\alpha$                & Hub-and-spoke intensity & S & --- \\
$\mathcal{D}^{dwell}$   & Dwell time (days); mean $\mu_D$ & S & Triangular \\
$\sigma_k$              & Arrival std.\ dev.\ (h) & S & --- \\
$CV^{TEU}_k$            & Per-call TEU CV & C & Gamma \\
$\mathcal{D}^{bp}_{ij}$ & Batch size per category pair & C & Gamma \\
Import/Export TEU       & Per-LSD cargo volume & --- & Gamma \\
Transship.\ discharge   & Per-LSD TS discharge & --- & Lognormal \\
Arrival slot width      & Inter-LSD time gap & --- & Gamma \\
\hline
\end{tabular}
\end{table}

\subsection{Phase 1: From Throughput to LSD Counts}
\label{sec:normalisation}

The central design goal of Phase~1 is to let the user specify a small
number of aggregate inputs---annual throughput $T$, category ratios $r_k$,
representative gross tonnages $\overline{GT}_k$, and transshipment
share~$\tau$---and have the framework derive a self-consistent vessel
schedule.  Per-LSD throughput is set proportional to GT, reflecting the
physical constraint that larger vessels carry more cargo:
$q_k \propto \overline{GT}_k$.
The transshipment share~$\tau$ decomposes~$q_k$
into cargo types: $\EE[Q^{td}_k] = (\tau/2)\,q_k$,\;
$\EE[Q^{imp}_k] = \EE[Q^{exp}_k] = ((1-\tau)/2)\,q_k$.
The mean weekly TEU \emph{handled} per LSD is therefore:
\begin{equation}
  \bar{q}_k = \EE[Q^{imp}_k] + \EE[Q^{exp}_k] + 2\,\EE[Q^{td}_k]
            = (1 + \tau)\,q_k,
  \label{eq:qbar}
\end{equation}
where the factor $(1+\tau)$ arises because each transshipment TEU is
handled twice (discharge and load).  The number of LSDs per category
then follows:
\begin{equation}
  N_k = \mathrm{round}\!\left(
    \frac{T_w}{\sum_{k'} r_{k'}\,\bar{q}_{k'}} \cdot r_k
  \right)\!,
  \quad T_w = \frac{7\,T}{365}.
  \label{eq:Nk}
\end{equation}
When the observed weekly call count~$N$ is available from port data,
the GT-proportional assumption yields the per-LSD throughput in
closed form:
\begin{equation}
  q_k = \frac{T_w \cdot \overline{GT}_k}
    {N\,(1+\tau)\,\sum_{k'} r_{k'}\,\overline{GT}_{k'}},
  \label{eq:qk_gt}
\end{equation}
so that all per-LSD TEU values are determined directly from the
five public inputs $(T,\,\tau,\,r_k,\,\overline{GT}_k,\,N)$ with no
separate scaling factor.  The cross-validation in
Section~\ref{sec:calibration} confirms $<$\,1\% error on real data.
If rounding yields $N_k = 0$ for any active category, that
category is merged with its nearest neighbour.

Finally, to ensure port-wide transshipment conservation ($\sum_k N_k
\EE[Q^{td}_k] = \sum_k N_k \EE[Q^{tl}_k]$), the mean transshipment \emph{loading} volume $\EE[Q^{tl}_j]$ for
category~$j$ is derived as:
\begin{equation}
  \EE[Q^{tl}_j] = \frac{1}{N_j}
    \sum_{i} \EE[Q^{td}_i] \cdot N_i \cdot \rho_{ij},
  \label{eq:transload}
\end{equation}
i.e., the total TEU discharged from all categories~$i$ and routed to
category~$j$ (via the OD ratios~$\rho_{ij}$), divided equally among the
$N_j$ LSDs of that category.

\subsection{Phase 2: LSD Initialisation and Flow Allocation}
\label{sec:lsd_init}

Phase~1 determines \emph{how many} LSDs exist in each category; Phase~2
gives each LSD an identity---a scheduled arrival time, a vessel size, and
a cargo profile---and then connects LSD pairs through transshipment flows.

\subsubsection{Attribute Sampling}

For each LSD $l = 1, \ldots, N_k$ in category $k$, the framework samples
a vessel length $\mathrm{LOA}_l \sim \mathcal{D}^{LOA}_k$ and expected
weekly TEU volumes $\bar{q}^{imp}_l$, $\bar{q}^{exp}_l$, and
$\bar{q}^{td}_l$ from category-specific distributions
(Table~\ref{tab:notation}).  The $N = \sum_k N_k$ LSDs must share a seven-day cycle.  Inter-arrival
slot widths $w_l$ are sampled from a Gamma distribution and rescaled to
sum to exactly seven days; each LSD's scheduled arrival is placed at the
slot midpoint $\tau_l = \sum_{l'<l} w_{l'} + w_l/2$.  After sampling, TEU
values are rescaled by a multiplicative correction factor to match category
means exactly:
\begin{equation}
  \bar{q}^x_l \leftarrow \bar{q}^x_l \cdot
    \frac{\EE[Q^x_k]}{N_k^{-1}\sum_{l'} \bar{q}^x_{l'}},
  \quad x \in \{imp, exp, td, tl\},
  \label{eq:normteu}
\end{equation}
where the numerator is the category-level mean from Phase~1 and the
denominator is the current sample mean across the $N_k$ LSDs of
category~$k$.  Sampling first, then correcting, keeps the inter-LSD heterogeneity
while forcing each category's aggregate throughput to match the Phase~1
targets.

\subsubsection{GreedyDwellFit Transshipment Allocation}

Each LSD now has a cargo profile but no inter-LSD connections.
For each category pair $(i,j)$, GDF (Algorithm~\ref{alg:gdf}) generates
TEU batches from $\mathcal{D}^{bp}_{ij}$ totalling $F_{ij} =
\EE[Q^{td}_i] \cdot N_i \cdot \rho_{ij}$, then processes them in
decreasing size---analogous to First~Fit~Decreasing---assigning each to
the LSD pair whose circular dwell $d_{l_1 l_2} = (\tau_{l_2} -
\tau_{l_1}) \bmod 7$ best matches a sampled target, subject to residual
capacity.  An adaptive soft threshold $\theta = e^{-f/10\,000}$
progressively relaxes constraints on failure; destination selection
favours unconnected pairs to promote OD diversity.

\begin{algorithm}[htb]
\caption{GreedyDwellFit (GDF) transshipment allocation.}
\label{alg:gdf}
\Input{LSD set $\calL$; residual discharge capacities $\{c^{td}_l\}$
  and loading capacities $\{c^{tl}_l\}$ (initialised to $\bar{q}^{td}_l$
  and $\bar{q}^{tl}_l$); flow pool $\calF$ sorted by batch size $b$
  descending; dwell distribution $\mathcal{D}^{dwell}$}
\Output{Flow matrix $F = \{f_{l_1 l_2}\}$; failure count $f^*$}
$\theta \leftarrow 1$; $f \leftarrow 0$\;
\While{$\calF \neq \emptyset$}{
  $(k_1, k_2, b) \leftarrow \mathrm{head}(\calF)$ \;
  $d \sim \mathcal{D}^{dwell}$ \Comment*[r]{sample target dwell}
  $\calL_1 \leftarrow \{l \in \calL : \mathrm{cat}(l)=k_1,\;
     c^{td}_l \geq \theta\, b\}$\;
  \If{$\calL_1 = \emptyset$}{
    $f \leftarrow f+1$; $\theta \leftarrow e^{-f/10\,000}$;
    \textbf{continue} \Comment*[r]{retry same batch}
  }
  $l_1 \sim \mathrm{Uniform}(\calL_1)$\;
  $\calL_2 \leftarrow \{l \in \calL : \mathrm{cat}(l)=k_2,\;
     c^{tl}_l \geq \theta\, b\}$\;
  \If{$\calL_2 = \emptyset$}{
    $f \leftarrow f+1$; $\theta \leftarrow e^{-f/10\,000}$;
    \textbf{continue}\;
  }
  $\calL_2^* \leftarrow \{l \in \calL_2 :
    l \notin \mathrm{dest}(l_1)\}$\;
  \lIf{$\calL_2^* \neq \emptyset$}{$\calL_2 \leftarrow \calL_2^*$}
  $l_2 \leftarrow \arg\min_{l \in \calL_2}
    \min_{j \in \{0,1,2\}}\bigl|(\tau_{l_1}+d)
    -\tau_l - 7j\bigr|$\;
  $f_{l_1 l_2} \leftarrow f_{l_1 l_2} + b$;
  $c^{td}_{l_1} \leftarrow c^{td}_{l_1} - b$;
  $c^{tl}_{l_2} \leftarrow c^{tl}_{l_2} - b$\;
  Remove head$(\calF)$\;
}
$f^* \leftarrow f$\;
\Return $F,\,f^*$
\end{algorithm}

The threshold $\theta$ decays monotonically toward zero on each failure,
so the capacity check $c^{td}_l \geq \theta\,b$ is eventually satisfied
by any LSD with positive residual capacity.  Since each successful
allocation removes a batch from~$\calF$, the pool size strictly decreases
and the algorithm terminates.  

\subsection{Phase 3: Stochastic Port Call Generation}
\label{sec:runtime}

Phases~1--2 produce a deterministic weekly template.  Phase~3 introduces
controlled stochasticity at two levels \shortcite{vanAsperen2003arrival}.
\emph{Level~1 (structural):} each LSD retains the distinct
$\tau_l$, $\bar{q}^x_l$, and $\mathrm{LOA}_l$ from Phase~2, fixed
for the simulation run.
\emph{Level~2 (operational):} at each weekly call, the actual time of
arrival~(ATA) is resampled:
\begin{equation}
  \mathrm{ATA}_l \sim \mathrm{Gamma}\!\bigl(\mu = \tau_l,\;
    CV = \sigma_k / (\tau_l \cdot 24)\bigr),
  \label{eq:ata}
\end{equation}
where $\tau_l$ is the LSD's scheduled arrival time (in days from the
start of the weekly cycle), $\tau_l \cdot 24$ converts it to hours, and
$\sigma_k$~(hours) is a category-specific arrival standard deviation.
The Gamma is parameterised with shape $\alpha = CV^{-2}$ and scale
$\beta = \mu \cdot CV^2$ \shortcite{law2015simulation}.

Because this Gamma's mean equals the scheduled arrival time, timing is
unbiased; its right skew is intentional, placing the heavy tail on
\emph{late} arrivals (rare large delays) while the positive support rules
out impossible early ones.

The Gamma is preferred over the Normal (which requires truncation to
avoid negative arrival times) and the Lognormal (excessive tail weight at
small CV).  Its shape parameter controls skewness: for large $\alpha$
(mainliners) it approximates the Normal, while for $\alpha \approx 1$
(feeders) it captures heavier tails matching empirical observations
\shortcite{rodrigues2016analysis}.

Per-call TEU volumes are similarly resampled: for each port call~$p$ of
LSD~$l$, $Q^x_{l,p} \sim \mathrm{Gamma}(\bar{q}^x_l,\, CV^{TEU}_k)$
for $x \in \{imp, exp, td\}$, with transshipment TEUs distributed across
destinations proportionally to the flow matrix entries~$f_{l l'}$.
Because each draw has mean $\bar{q}^x_l$, the mean throughput equals~$T$
by construction, with no runtime rescaling; a finite run's realised total
only approximates~$T$ (empirical error $e_T < 1.5\%$,
Section~\ref{sec:exp4}).
This strict separation of the unbiased global mean ($T$) from the
user-controlled local variance ($CV^{TEU}$) prevents random volume overshoots 
from triggering artificial, non-linear queueing congestion, 
enabling rigorous ``apples-to-apples'' scenario comparisons.

The framework computes built-in validation metrics at generation time.
For each LSD~$l$, the normalised discharge residual
$\varepsilon^{td}_l = c^{td}_l / \bar{q}^{td}_l$ measures
how much of the expected discharge capacity was left unallocated by
GDF, and likewise $\varepsilon^{tl}_l$ for loading.  Averaging the
absolute values across all LSDs gives $\mathrm{MAE}^{td}$ and
$\mathrm{MAE}^{tl}$.  A global dwell time error
$\mathrm{MAE}^{dwell} = |\bar{D} - \mu_D|$ compares the realised mean
dwell (across all allocated flows) against the target mean~$\mu_D$.
These can be checked before running the simulation.

\section{COMPUTATIONAL EXPERIMENTS}
\label{sec:experiments}

We test the framework in four experiments: allocation quality
(Experiment~1), sensitivity to arrival-process and transshipment-model
choices (Experiments~2--3), and cross-validation against Singapore with
throughput scaling and sensitivity analysis (Experiment~4).

\subsection{Setup}
\label{sec:setup}

\subsubsection{Base Configuration}

All experiments use the Port of Busan~\shortcite{bpa2024stats}
(23.15\,M~TEU, $\tau = 54.2\%$) as the primary configuration.
Table~\ref{tab:base_config} lists the five public inputs and per-LSD
throughput from Equation~(\ref{eq:qk_gt}).  Category ratios are inferred
from AIS vessel-size analyses~\shortcite{unctad2024maritime}: 70\% of calls
involve ships under 4\,000~TEU, with the rest split between mainliners
(18\%) and ULCVs (12\%).  Proprietary data (OD matrices, per-service
manifests) can replace any default.

\begin{table}[htb]
\centering
\caption{Port configurations.  Cat~1 = feeder (46--147\,m);
  Cat~2 = regional (147--209\,m); Cat~3 = mainliner (209--285\,m);
  Cat~4 = ULCV (285--400\,m).  Per-LSD throughput $q_k$ derived from
  Equation~(\ref{eq:qk_gt}).  Busan is the primary test bed;
  Singapore is used for cross-validation
  (Section~\ref{sec:calibration}).}
\label{tab:base_config}
\label{tab:calibrated}
\footnotesize
\setlength{\tabcolsep}{3pt}
\begin{tabular}{lrrrr|rrrr}
\hline
 & \multicolumn{4}{c|}{\textbf{Busan}} & \multicolumn{4}{c}{\textbf{Singapore}} \\
Parameter & Cat~1 & Cat~2 & Cat~3 & Cat~4
          & Cat~1 & Cat~2 & Cat~3 & Cat~4 \\
\hline
LOA range (m)          & 46--147 & 147--209 & 209--285 & 285--400
                       & \multicolumn{4}{c}{(same)} \\
Mean LOA (m)           & 90  & 176  & 250 & 338
                       & \multicolumn{4}{c}{(same)} \\
$\overline{GT}_k$      & 5\,000 & 25\,000 & 65\,000 & 180\,000
                       & \multicolumn{4}{c}{(same)} \\
$r_k$                  & 0.35 & 0.35 & 0.18 & 0.12
                       & 0.306 & 0.233 & 0.259 & 0.202 \\
$q_k$ (TEU/LSD)        & 164  & 821  & 2\,136 & 5\,912
                       & 128  & 637  & 1\,657 & 4\,588 \\
$\bar{q}_k$ (TEU/LSD)  & 253  & 1\,266 & 3\,294 & 9\,127
                       & 236  & 1\,179 & 3\,065 & 8\,487 \\
$N_k$ (LSDs/week)      & 70   & 70   & 36  & 24
                       & 92   & 70   & 78  & 61 \\
$CV^{TEU}$             & 0.3  & 0.3  & 0.3 & 0.3
                       & \multicolumn{4}{c}{(same)} \\
$\sigma_{arrival}$ (h) & 3    & 3    & 3   & 3
                       & \multicolumn{4}{c}{(same)} \\
\hline
\multicolumn{9}{l}{\footnotesize Busan: $T = 23.15$\,M, $\tau = 0.542$,
  $N = 200$/wk, $\alpha = 0.5$} \\
\multicolumn{9}{l}{\footnotesize Singapore: $T = 44.66$\,M, $\tau = 0.85$,
  $N = 300$/wk, $\alpha = 0.5$} \\
\end{tabular}
\end{table}

\paragraph{Transshipment ratio matrix.}
The $4 \times 4$ OD ratio matrix $\{\rho_{ij}\}$ specifies what fraction
of category-$i$ discharge TEUs are loaded onto category-$j$ vessels.
Rather than prescribing 12~free parameters (16~cells minus 4~row-sum
constraints) without empirical support, we adopt a single-parameter
\emph{gravity model}:
\begin{equation}
  \rho_{ij} = \frac{r_j \,\exp\!\bigl(\alpha\,|s_i - s_j|\bigr)}
    {\sum_{k} r_k \,\exp\!\bigl(\alpha\,|s_i - s_k|\bigr)},
  \label{eq:gravity}
\end{equation}
where $s_k = \overline{GT}_k / \max_k \overline{GT}_k \in [0,1]$ is
a normalised vessel-size index and $\alpha$ is the \emph{hub-and-spoke
intensity}:
(i)~$\alpha = 0$ gives a \emph{capacity-proportional} baseline where
cargo flows to each category in proportion to~$r_j$;
(ii)~$\alpha > 0$ produces a \emph{hub-and-spoke} pattern where
unlike-size categories exchange more
(feeders~$\leftrightarrow$~ULCVs);
(iii)~$\alpha < 0$ favours \emph{intra-tier} exchange, as in a gateway
port with direct deep-sea services.
Note that, unlike a classical gravity model where interaction decreases
with distance, here the exponential \emph{increases} with size
difference, reflecting the hub-and-spoke pattern in which feeders and
mainliners exchange disproportionately.
This reduces the OD specification from 12~unjustifiable parameters to a
single interpretable one.  The default configuration uses $\alpha = 0.5$
(moderate hub-and-spoke), reflecting the typical pattern at transshipment
hubs where feeder--mainliner exchange dominates; the sensitivity of
simulation outputs to~$\alpha$ is quantified in the OAT analysis
(Section~\ref{sec:exp4}).  If a port
planner has access to real OD flow data (e.g., from customs manifests or
terminal operating systems), the matrix can be replaced directly,
bypassing the gravity model entirely.

Transshipment dwell time follows a Triangular(1,\,4,\,7) distribution (days).

\subsubsection{Simulation Protocol}

The schedule generator runs on the O2DESNet discrete-event
engine~\shortcite{li2015object}, which advances a simulation clock and fires
stochastic port call events over 56 simulated weeks (8-week warm-up,
determined by inspecting convergence of cumulative-average plots
\shortcite{law2015simulation}), with 30~replications per condition.  The
generator exposes each port call through a callback interface, so its
output can equally feed any external DES framework.  Experiments~2--3 use
lightweight analytical test beds---a continuous-berth queuing model and a
yard inventory bookkeeping model---that consume the generated port calls
to isolate the effect of the arrival process and transshipment structure,
respectively; they are not intended as full-fidelity terminal simulators.
Common random numbers~(CRN) are used for variance reduction in
Experiments~2--3: all arrival models (or transshipment models) share the
same LSD template and seed per replication, so that paired differences
isolate the effect of the model choice.  Full synchronisation is achieved
by assigning dedicated random streams to arrival-time sampling and
TEU-volume sampling, so that differences in the number of draws consumed
by each arrival model cannot shift the cargo stream
\shortcite{law2015simulation}.  Pairwise comparisons are tested via
paired-$t$ 95\% confidence intervals; an interval excluding zero
indicates a statistically significant difference at $\alpha = 0.05$.

\subsection{Experiment 1: GDF Allocation Quality}
\label{sec:exp1}

\emph{Research question}: Does the decreasing-size processing order of
GDF produce a better OD flow allocation than random-order assignment?

GDF is compared against TotalRandom, which processes batches in
random order with slower threshold decay ($\theta = e^{-f/30\,000}$).
One hundred independent schedules are generated per variant using the
Busan configuration (Table~\ref{tab:base_config}).

A synthetic recovery test checks whether the generated schedules
preserve the input OD ratios~$\rho_{ij}$.  Both allocators achieve
Pearson $r = 1.000$ across 100~schedules, so the category-level flow
structure is intact.  Where GDF wins is at the per-LSD level: it cuts
discharge error by two orders of magnitude ($0.02\%$ vs.\ $1.71\%$) and
reduces failures by 38$\times$, with 58\% shorter run times because fewer
retries are needed.  Dwell-time error is similar for both
($\approx$\,4\%)---a hard floor set by the mismatch between a continuous
dwell target and discrete weekly slots.

\begin{table}[htb]
\centering
\caption{Schedule generation quality (mean $\pm$ std.\ dev., $n=100$).}
\label{tab:e1}
\footnotesize
\begin{tabular}{lrr}
\hline
Metric & TotalRandom & GDF \\
\hline
$\mathrm{MAE}^{td}$ (\%)
  & $1.71\pm0.38$ & $\mathbf{0.02\pm0.01}$ \\
$\mathrm{MAE}^{tl}$ (\%)
  & $0.43\pm0.07$ & $\mathbf{0.00\pm0.00}$ \\
$\mathrm{MAE}^{dwell}$ (\%)
  & $4.43\pm1.41$ & $3.93\pm1.23$ \\
Failure count $f^*$
  & $183\,979\pm50\,355$ & $\mathbf{4\,862\pm1\,336}$ \\
CPU time (s)
  & $1.20\pm0.18$ & $\mathbf{0.50\pm0.07}$ \\
OD recovery $r$
  & $1.000\pm0.000$ & $1.000\pm0.000$ \\
\hline
\end{tabular}
\end{table}

\subsection{Experiment 2: Arrival Process Comparison}
\label{sec:exp2}

\emph{Research question}: How sensitive are terminal performance
predictions to the choice of arrival process model?

Four arrival models are compared using a continuous-berth queuing model.
Each vessel occupies $\mathrm{LOA} + 30$\,m of wharf frontage under FCFS
scheduling.  The number of quay cranes $n_c$ assigned to each vessel call
is drawn from a discretised triangular distribution conditioned on vessel
LOA and per-call TEU volume, implemented as a $4 \times 4$ category matrix
with mean $n_c$ ranging from 1.2 (smallest feeders) to 4.8 (largest
post-Panamaxes).  Service time is then
\begin{equation}
  s = \frac{Q}{n_c \cdot r_{qc} \cdot c_{\mathrm{TEU}}} + t_{\mathrm{setup}},
  \quad s \geq 4\;\text{h},
  \label{eq:service_time}
\end{equation}
where $Q$ is total TEU handled, $r_{qc} = 25$ moves per crane-hour,
$c_{\mathrm{TEU}} = 1.4$ TEU per move (reflecting the 40-ft container
ratio), and $t_{\mathrm{setup}} = 2.5$\,h is a fixed
berthing/unberthing overhead.  To expose the regime where arrival-model
choice matters most, we test two wharf-length settings: a
\emph{baseline} scenario ($L = 8\,000$\,m, moderate utilisation) and a
\emph{stressed} scenario ($L = 6\,500$\,m, high utilisation).

\begin{description}
  \item[M1 -- Poisson] Homogeneous Poisson process \shortcite{legato2001berth},
    $\lambda = N/7$ per day.
  \item[M2 -- Equidistant] Deterministic: each LSD fires at $\tau_l$
    every 7~days.
  \item[M3 -- Uniform window] ATA sampled uniformly within the
    inter-arrival slot.
  \item[M4 -- Proposed (Gamma)] Gamma-perturbed as in
    Equation~(\ref{eq:ata}).
\end{description}

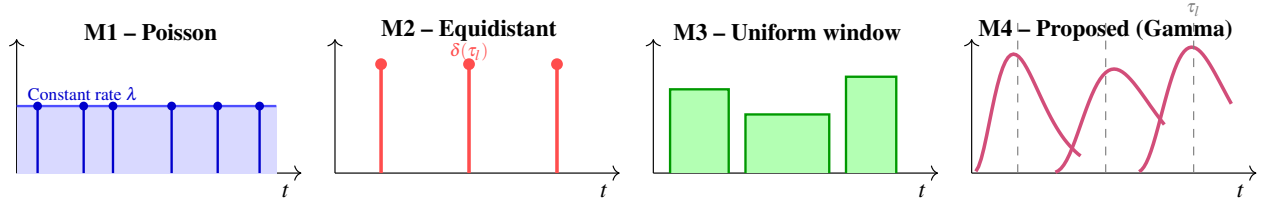
\begin{figure}[htb]
\centering
\resizebox{\columnwidth}{!}{%
\begin{tikzpicture}
\begin{scope}[local bounding box=plots]
\def\pw{3.2}
\def\ph{1.4}
\def\gap{0.6}
\begin{scope}[shift={(0,0)}]
 \draw[->] (0,0) -- (\pw+0.2,0) node[below left, font=\scriptsize]{$t$};
 \draw[->] (0,0) -- (0,\ph+0.2);
 \node[font=\scriptsize\bfseries, above] at (\pw/2, \ph+0.05) {M1 -- Poisson};
 \fill[blue!15] (0, 0) rectangle (3.1, 0.8);
 \draw[blue!70, thick] (0, 0.8) -- (3.1, 0.8);
 \node[font=\tiny, blue!80!black, right] at (0, 0.95) {Constant rate $\lambda$};
 \foreach \x in {0.25, 0.8, 1.15, 1.85, 2.4, 2.9} {
   \draw[blue!80!black, thick] (\x, 0) -- (\x, 0.8);
   \fill[blue!80!black] (\x, 0.8) circle (1.5pt);
 }
\end{scope}
\begin{scope}[shift={(\pw+\gap, 0)}]
 \draw[->] (0,0) -- (\pw+0.2,0) node[below left, font=\scriptsize]{$t$};
 \draw[->] (0,0) -- (0,\ph+0.2);
 \node[font=\scriptsize\bfseries, above] at (\pw/2, \ph+0.05) {M2 -- Equidistant};
 \foreach \x in {0.55, 1.60, 2.65} {
   \draw[red!70, very thick] (\x, 0) -- (\x, 1.3);
   \fill[red!70] (\x, 1.3) circle (2pt);
 }
 \node[font=\tiny, red!70] at (1.6, 1.45) {$\delta(\tau_l)$};
\end{scope}
\begin{scope}[shift={(2*\pw+2*\gap, 0)}]
 \draw[->] (0,0) -- (\pw+0.2,0) node[below left, font=\scriptsize]{$t$};
 \draw[->] (0,0) -- (0,\ph+0.2);
 \node[font=\scriptsize\bfseries, above] at (\pw/2, \ph+0.05) {M3 -- Uniform window};
 \fill[green!30] (0.2,0) rectangle (0.9, 1.0);
 \draw[green!60!black, thick] (0.2,0) -- (0.2,1.0) -- (0.9,1.0) -- (0.9,0);
 \fill[green!30] (1.1,0) rectangle (2.1, 0.7);
 \draw[green!60!black, thick] (1.1,0) -- (1.1,0.7) -- (2.1,0.7) -- (2.1,0);
 \fill[green!30] (2.3,0) rectangle (2.9, 1.15);
 \draw[green!60!black, thick] (2.3,0) -- (2.3,1.15) -- (2.9,1.15) -- (2.9,0);
\end{scope}
\begin{scope}[shift={(3*\pw+3*\gap, 0)}]
 \draw[->] (0,0) -- (\pw+0.2,0) node[below left, font=\scriptsize]{$t$};
 \draw[->] (0,0) -- (0,\ph+0.2);
 \node[font=\scriptsize\bfseries, above] at (\pw/2, \ph+0.05) {M4 -- Proposed (Gamma)};
 \draw[purple!70, very thick, domain=0.05:1.3, samples=40]
   plot (\x, {1.4*(\x/0.55)^3 * exp(-(\x-0.55)*6)});
 \draw[purple!70, very thick, domain=1.0:2.3, samples=40]
   plot (\x, {1.2*((\x-0.9)/0.7)^4 * exp(-(\x-1.6)*5)});
 \draw[purple!70, very thick, domain=2.0:3.1, samples=40]
   plot (\x, {1.5*((\x-1.9)/0.75)^4 * exp(-(\x-2.65)*5.5)});
 \foreach \x in {0.55, 1.60, 2.65} {
   \draw[gray, dashed, thin] (\x, 0) -- (\x, 1.85);
 }
 \node[font=\tiny, gray] at (2.65, 1.95) {$\tau_l$};
\end{scope}
\end{scope}
\end{tikzpicture}%
}
\caption{Schematic probability density of actual arrival times over the weekly cycle
 under the four models.}
\label{fig:interarrival}
\end{figure}

M1 and M2 represent the extremes from \shortciteN{vanAsperen2003arrival};
M3 is a common intermediate.  M4 should fall between M2 and M1 but
closer to M2.  Performance is measured by mean vessel waiting
time~$W_v$~(h), $\sigma_{W_v}$, and $P(W_v > 4\,\text{h})$.

Under the baseline scenario (Table~\ref{tab:e2}a), the system operates
at moderate utilisation ($\rho \approx 0.76$): M2, M3, and M4 all produce
near-zero waiting, while Poisson~(M1) inflates $W_v$ by over an order of
magnitude.  A paired-$t$ test confirms that the M1--M4 difference is
significant ($\Delta W_v = 0.31$\,h, 95\%~CI $[0.26,\, 0.36]$).  The
M2--M4 and M3--M4 differences are statistically significant but
practically negligible ($|\Delta W_v| < 0.01$\,h, i.e.\ under
30~seconds).

\begin{table}[htb]
\centering
\caption{Terminal performance under four arrival models
         (mean $\pm$ std.\ dev., $n=30$).  Panel~(a): baseline
         wharf ($L = 8\,000$\,m, $\rho \approx 0.76$).
         Panel~(b): stressed wharf ($L = 6\,500$\,m, $\rho \approx 0.94$).}
\label{tab:e2}
\footnotesize
\begin{tabular}{lrrrr}
\hline
\multicolumn{5}{l}{\emph{(a) Baseline ($L = 8\,000$\,m)}} \\
\hline
Metric & M1 & M2 & M3 & M4 \\
\hline
$W_v$ (h)              & $0.34\pm0.12$ & $0.02\pm0.01$ & $0.02\pm0.02$ & $\mathbf{0.03\pm0.02}$ \\
$\sigma_{W_v}$ (h)     & $1.40\pm0.42$ & $0.19\pm0.09$ & $0.20\pm0.10$ & $\mathbf{0.24\pm0.11}$ \\
$P(W_v>4\,\text{h})$   & $0.03\pm0.01$ & $0.00\pm0.00$ & $0.00\pm0.00$ & $\mathbf{0.00\pm0.00}$ \\
$\rho_{berth}$         & $0.76\pm0.01$ & $0.76\pm0.01$ & $0.76\pm0.01$ & $\mathbf{0.76\pm0.01}$ \\
\hline
\multicolumn{5}{l}{\emph{(b) Stressed ($L = 6\,500$\,m)}} \\
\hline
Metric & M1 & M2 & M3 & M4 \\
\hline
$W_v$ (h)              & $3.80\pm0.78$ & $1.59\pm0.39$ & $1.60\pm0.45$ & $\mathbf{1.72\pm0.44}$ \\
$\sigma_{W_v}$ (h)     & $6.35\pm1.17$ & $2.88\pm0.53$ & $2.88\pm0.69$ & $\mathbf{3.08\pm0.74}$ \\
$P(W_v>4\,\text{h})$   & $0.30\pm0.05$ & $0.15\pm0.04$ & $0.15\pm0.04$ & $\mathbf{0.17\pm0.04}$ \\
$\rho_{berth}$         & $0.93\pm0.01$ & $0.94\pm0.01$ & $0.94\pm0.01$ & $\mathbf{0.94\pm0.01}$ \\
\hline
\end{tabular}
\end{table}

The stressed scenario (Table~\ref{tab:e2}b) reduces the wharf to
6\,500\,m, pushing utilisation to $\rho \approx 0.94$ where queueing
effects become pronounced.  Poisson~(M1) now predicts
$W_v = 3.80$\,h---2.2$\times$ that of the Gamma model~(M4, $1.72$\,h).
The paired difference is $\Delta W_v = 2.08$\,h (95\%~CI
$[1.78,\, 2.39]$), and $\Delta P(W_v > 4\text{h}) = 0.14$ (95\%~CI
$[0.12,\, 0.16]$)---both highly significant.  By contrast, the M2--M4
and M3--M4 differences are not significant (CIs include zero),
confirming that for liner traffic the primary distinction is between
memoryless (M1) and schedule-aware models (M2--M4).  M4 sits between M2
and M1 but much closer to M2, consistent with the observation that liner
arrivals are structured with modest perturbation.

These results show that restoring cyclic schedule structure eliminates
the artificial congestion created by Poisson arrivals.  The practical
advantage of M4 over M2 and M3 lies not in the mean but in the tails:
M2 assumes zero arrival variance and M3 imposes rigid uniform bounds,
whereas the Gamma distribution in M4 offers a tunable shape parameter
that can represent the unbounded, heavy-tailed delays characteristic of
maritime disruptions (port congestion, weather, canal queuing).  This
makes M4 the appropriate choice when stress-testing terminal capacity
under extreme peak scenarios.

\subsection{Experiment 3: Transshipment Flow Model Comparison}
\label{sec:exp3}

\emph{Research question}: Does an explicit OD transshipment structure
produce meaningfully different yard dynamics than aggregate or
no-transshipment models?

Using arrival model M4 and a fixed LSD schedule, three levels of
transshipment modelling fidelity are compared:

\begin{description}
  \item[T1 -- No transshipment] All cargo is import/export only.
  \item[T2 -- Aggregate] Transshipment TEUs are generated per call, but
    destination LSDs are selected uniformly at random.
  \item[T3 -- Proposed (OD matrix)] Destinations determined by the
    GDF flow matrix.
\end{description}

A yard inventory bookkeeping model isolates the OD effect from crane
and vehicle confounders: import containers dwell
$\mathrm{Triangular}(1,\,2,\,4)$~days; export containers arrive
$6 \times U(0.5,\,1.5)$~days before their vessel; transshipment
containers exit at the next call of their destination LSD.
Yard occupancy is sampled hourly.  Key metrics are mean yard
occupancy~$\bar{Y}$, realised mean transshipment dwell
time~$\bar{D}_{actual}$, and its standard deviation~$\sigma_D$.

Without transshipment (T1), yard occupancy drops by 34\%, confirming that
transshipment containers dominate yard inventory.  Comparing T2 and T3,
the OD model raises occupancy by 2\,766~containers
($\Delta\bar{Y}$ 95\%~CI $[2\,620,\, 2\,913]$, $p < 0.05$)
because it bunches container departures around specific days rather than
spreading them evenly.  The more telling difference is in dwell-time
variability: $\sigma_D$ drops from 2.03 to 1.56~days under T3
($\Delta\sigma_D = 0.47$~days, 95\%~CI $[0.45,\, 0.48]$), because
containers paired to specific outbound services have narrower departure
windows than randomly assigned ones.  All paired differences are
statistically significant at the 95\% level.

\begin{table}[htb]
\centering
\caption{Yard and terminal performance under three transshipment models
         (mean $\pm$ std.\ dev., $n=30$).}
\label{tab:e3}
\footnotesize
\begin{tabular}{lrrr}
\hline
Metric & T1 & T2 & T3 \\
\hline
$\bar{Y}$ (containers)   & $64\,056\pm296$ & $93\,867\pm336$ & $\mathbf{96\,634\pm532}$ \\
$\bar{D}_{actual}$ (days) & ---             & $3.51\pm0.01$ & $\mathbf{3.84\pm0.05}$ \\
$\sigma_D$ (days)         & ---             & $2.03\pm0.00$ & $\mathbf{1.56\pm0.04}$ \\
\hline
\end{tabular}
\end{table}

\subsection{Experiment 4: Cross-Validation and Scaling}
\label{sec:exp4}

Experiment~4 checks the framework against real port data and tests whether
the normalisation holds across different throughput levels.

\subsubsection{Cross-Validation Against Singapore}
\label{sec:calibration}

The five-input procedure is applied to Singapore using 2025 MPA
statistics~\shortcite{mpa2025stats}.  Singapore's 85\% transshipment
share---far higher than Busan's 54\%---makes it an ideal
cross-validation case.  MPA data report 15\,607 container vessel calls
in 2025 with an average GT of 61\,555; this constrains the fleet-size
mix via interpolation, yielding $r = [0.306,\, 0.233,\, 0.259,\, 0.202]$
and reproducing the observed GT exactly.  Singapore's published
$\tau = 85\%$ decomposes each category's per-LSD TEU into
$\EE[Q^{td}_k] = 0.425\,q_k$ and
$\EE[Q^{imp}_k] = \EE[Q^{exp}_k] = 0.075\,q_k$.
With $N = 15\,607 / 52 \approx 300$ calls per week,
Equation~(\ref{eq:qk_gt}) yields the Singapore columns of
Table~\ref{tab:base_config}, with no separate scaling factor required.

The structural differences between the two ports are consistent:
Singapore's higher ULCV share ($r_4 = 0.202$ vs.\ $0.12$) and much
higher $\tau$ ($0.85$ vs.\ $0.542$) reflect its mega-hub role, while
Busan's feeder-heavy fleet is typical of a regional relay hub.
Both ports' call counts and annual throughput are matched to within
0.5\%, supporting the framework's generalisability without ad-hoc
scaling.

\subsubsection{Throughput Scaling}

To verify that the normalisation generalises beyond a single throughput
level, the framework is instantiated at five levels from $T = 10$\,M to
$50$\,M~TEU/yr ($n=30$ replications each, $T_0 = 30$\,M).
LSD counts $N = [86, 172, 260, 345, 432]$ scale linearly with
$T/T_0$ ($R^2 > 0.999$), matching the theoretical ratios $N/N_0 =
[0.33, 0.66, 1.00, 1.33, 1.66]$ against $T/T_0 = [0.33, 0.67, 1.00,
1.33, 1.67]$.  The relative throughput error $e_T = |T_{realised} - T|/T$
stays below 1.5\% at all levels (max $1.42\pm0.37$\% at 10\,M), confirming
that the normalisation works across the full range of major global port
sizes.

\subsubsection{Sensitivity Analysis}

A one-at-a-time~(OAT) sensitivity analysis over six parameters
identifies which inputs have the greatest leverage on~$W_v$ and~$\bar{Y}$.
Normalised indices $S(y,x) = (\Delta y / \bar{y}) / (\Delta x /
\bar{x})$ are reported; OAT does not capture interactions.

The two outputs respond to different inputs.  Waiting time~$W_v$ is
dominated by the transshipment percentage ($|S| = 15.82$): a 50\% volume
increase pushes berth utilisation into the regime where queuing delays
grow non-linearly.  Category mix is a distant second
($|S| = 0.63$), followed by arrival variability ($|S| = 0.41$), in line
with Experiment~2.  Hub-and-spoke intensity~$\alpha$ has negligible
effect ($|S| = 0.04$), indicating that the OD routing structure
redistributes cargo among LSDs but does not change total berth demand.
Yard occupancy~$\bar{Y}$ responds mainly to
transshipment volume ($|S| = 0.22$) and dwell-time mean
($|S| = 0.21$), since both directly govern how many containers sit in
the yard.  TEU variability, arrival variability, category mix, and
$\alpha$ have negligible effect on yard occupancy.

\begin{table}[htb]
\centering
\caption{OAT sensitivity indices $|S(y,x)|$ for vessel waiting
  time $W_v$ and mean yard occupancy $\bar{Y}$.  Each parameter is
  varied between the low and high values shown while all others are
  held at baseline ($n=30$ replications per setting).}
\label{tab:sensitivity}
\footnotesize
\begin{tabular}{lccc|cc}
\hline
Parameter & Low & High & Baseline & $|S(W_v,x)|$ & $|S(\bar{Y},x)|$ \\
\hline
TS\%                  & 0.5  & 1.5 & 1.0 & \textbf{15.82} & \textbf{0.22} \\
$\sigma_{arrival}$ (h) & 1   & 9   & 3   & 0.41           & $<$0.01 \\
$\mu_D$ (days)        & 3    & 5   & 4   & 0.08           & 0.21 \\
$CV^{TEU}$            & 0.1  & 0.5 & 0.3 & 0.04           & $<$0.01 \\
Cat.\ mix shift       & $-$0.5 & 0.5 & 0 & 0.63           & $<$0.01 \\
$\alpha$ (hub intensity) & 0  & 1.0  & 0.5 & 0.04         & $<$0.01 \\
\hline
\end{tabular}
\end{table}

\paragraph{Implications for what-if capacity studies.}
Because queueing systems degrade non-linearly near capacity, even minor
deviations from the target~($T$) distort what-if capacity studies (e.g.\
a guaranteed Berthing-On-Arrival rate).  Fixing the expected throughput
at~$T$ (realised within about 1\%, Section~\ref{sec:exp4}) while $CV^{TEU}$
tunes arrival volatility independently enables clean, apples-to-apples
scenario comparison without confounding overall demand.

\section{CONCLUSIONS}
\label{sec:conclusion}

High-fidelity transshipment hub simulation has long been limited by simplistic Poisson arrivals and the unavailability of proprietary schedules, which we overcome with a generator that derives structurally realistic vessel traffic from five public, macroscopic statistics. It couples a two-level Gamma arrival model---mirroring the cyclic nature of liner shipping while allowing operational delays---with GreedyDwellFit (GDF), a fast heuristic that pairs transshipment cargo to specific connecting services; together these prevent the artificial inflation of vessel queues and yield more realistic yard-dwell patterns than legacy aggregate models.

Because the generator is data-agnostic, planners can immediately deploy it without proprietary manifests. Future work should validate against per-service schedule data, apply factorial designs to capture parameter interactions, and integrate the generator into full-scale terminal models to assess its impact on berth allocation and crane productivity.

\section*{ACKNOWLEDGMENTS}

The authors gratefully acknowledge the funding support from the 
Singapore Maritime Institute under project grant ID: SMI-2022-SP-02,
which made this research possible.

\bibliographystyle{wsc}
\bibliography{references}

\begin{thebibliography}{}

\bibitem[\protect\citeauthoryear{Bierwirth and Meisel}{Bierwirth and
  Meisel}{2015}]{bierwirth2015survey}
Bierwirth, C., and F.~Meisel. 2015.
\newblock ``A Follow-Up Survey of Berth Allocation and Quay Crane Scheduling
  Problems in Container Terminals''.
\newblock {\em European Journal of Operational
  Research\/}~244(3):675--689~\url{https://doi.org/10.1016/j.ejor.2014.12.030}.


\bibitem[\protect\citeauthoryear{Brouer, Karsten, and Pisinger}{Brouer
  et~al.}{2017}]{brouer2017vessel}
Brouer, B.~D., C.~V. Karsten, and D.~Pisinger. 2017.
\newblock ``Optimization in Liner Shipping''.
\newblock {\em 4OR -- A Quarterly Journal of Operations
  Research\/}~15(1):1--35~\url{https://doi.org/10.1007/s10288-017-0342-6}.


\bibitem[\protect\citeauthoryear{{Busan Port Authority}}{{Busan Port
  Authority}}{2024}]{bpa2024stats}
{Busan Port Authority} 2024.
\newblock ``Container Throughput Statistics''.
\newblock \url{https://www.busanpa.com/eng/Contents.do?mCode=MN0042}.
\newblock Accessed: 2026-04-01. Annual container throughput and
  import/export/transshipment breakdown.

\bibitem[\protect\citeauthoryear{Canonaco, Legato, Mazza, and
  Musmanno}{Canonaco et~al.}{2008}]{canonaco2008queuing}
Canonaco, P., P.~Legato, R.~M. Mazza, and R.~Musmanno. 2008.
\newblock ``A Queuing Network Model for the Management of Berth Crane
  Operations''.
\newblock {\em Computers \& Operations
  Research\/}~35(8):2432--2446~\url{https://doi.org/10.1016/j.cor.2006.12.001}.


\bibitem[\protect\citeauthoryear{Di~Crescenzo, Martinucci, and
  Paraggio}{Di~Crescenzo et~al.}{2023}]{dicrescenzo2023vessels}
Di~Crescenzo, A., B.~Martinucci, and P.~Paraggio. 2023.
\newblock ``Vessels Arrival Process and its Application to the
  {SHIP/M/$\infty$} Queue''.
\newblock {\em Methodology and Computing in Applied
  Probability\/}~25(1):1--33~\url{https://doi.org/10.1007/s11009-023-10003-8}.


\bibitem[\protect\citeauthoryear{Dragovi\'{c}, Tzannatos, and
  Park}{Dragovi\'{c} et~al.}{2017}]{dragovic2017simulation}
Dragovi\'{c}, B., E.~Tzannatos, and N.~K. Park. 2017.
\newblock ``Simulation Modelling in Ports and Container Terminals: Literature
  Overview and Analysis by Research Field, Application Area and Tool''.
\newblock {\em Flexible Services and Manufacturing
  Journal\/}~29(1):4--34~\url{https://doi.org/10.1007/s10696-016-9239-5}.


\bibitem[\protect\citeauthoryear{Hartmann}{Hartmann}{2004}]{hartmann2004generating}
Hartmann, S. 2004.
\newblock ``Generating Scenarios for Simulation and Optimization of Container
  Terminal Logistics''.
\newblock {\em OR
  Spectrum\/}~26(2):171--192~\url{https://doi.org/10.1007/s00291-003-0150-6}.


\bibitem[\protect\citeauthoryear{Hendriks, Laumanns, Lefeber, and
  Udding}{Hendriks et~al.}{2010}]{hendriks2010robust}
Hendriks, M., M.~Laumanns, E.~Lefeber, and J.~T. Udding. 2010.
\newblock ``Robust Cyclic Berth Planning of Container Vessels''.
\newblock {\em OR
  Spectrum\/}~32(3):501--517~\url{https://doi.org/10.1007/s00291-010-0198-z}.


\bibitem[\protect\citeauthoryear{Jia, Li, and Xu}{Jia
  et~al.}{2020}]{jia2020simulation}
Jia, S., C.-L. Li, and Z.~Xu. 2020.
\newblock ``A Simulation Optimization Method for Deep-Sea Vessel Berth Planning
  and Feeder Arrival Scheduling at a Container Port''.
\newblock {\em Transportation Research Part B:
  Methodological\/}~142:174--196~\url{https://doi.org/10.1016/j.trb.2020.10.007}.


\bibitem[\protect\citeauthoryear{Law}{Law}{2015}]{law2015simulation}
Law, A.~M. 2015.
\newblock {\em Simulation Modeling and Analysis\/}. 5th ed.
\newblock New York: McGraw-Hill Education.


\bibitem[\protect\citeauthoryear{Legato and Mazza}{Legato and
  Mazza}{2001}]{legato2001berth}
Legato, P., and R.~M. Mazza. 2001.
\newblock ``Berth Planning and Resources Optimisation at a Container Terminal
  via Discrete Event Simulation''.
\newblock {\em European Journal of Operational
  Research\/}~133(3):537--547~\url{https://doi.org/10.1016/S0377-2217(00)00200-9}.


\bibitem[\protect\citeauthoryear{Li, Zhu, Chen, Pedrielli, and Pujowidianto}{Li
  et~al.}{2015}]{li2015object}
Li, H., Y.~Zhu, Y.~Chen, G.~Pedrielli, and N.~A. Pujowidianto. 2015.
\newblock ``The Object-Oriented Discrete Event Simulation Modeling: A Case
  Study on Aircraft Spare Part Management''.
\newblock In {\em 2015 Winter Simulation Conference (WSC)},
  3514--3525~\url{https://doi.org/10.1109/WSC.2015.7408511}.

\bibitem[\protect\citeauthoryear{{Maritime and Port Authority of
  Singapore}}{{Maritime and Port Authority of Singapore}}{2026}]{mpa2025stats}
{Maritime and Port Authority of Singapore} 2026.
\newblock ``Port Statistics''.
\newblock
  \url{https://www.mpa.gov.sg/who-we-are/newsroom-resources/research-and-statistics/port-statistics}.
\newblock Accessed: 2026-07-29. Public statistics on container throughput,
  vessel arrivals, and vessel calls.

\bibitem[\protect\citeauthoryear{Petering, Wu, Li, Goh, and de~Souza}{Petering
  et~al.}{2009}]{petering2009development}
Petering, M. E.~H., Y.~Wu, W.~Li, M.~Goh, and R.~de~Souza. 2009.
\newblock ``Development and Simulation Analysis of Real-Time Yard Crane Control
  Systems for Seaport Container Transshipment Terminals''.
\newblock {\em OR
  Spectrum\/}~31(4):801--835~\url{https://doi.org/10.1007/s00291-008-0142-7}.


\bibitem[\protect\citeauthoryear{Rodrigues and Rangel}{Rodrigues and
  Rangel}{2016}]{rodrigues2016analysis}
Rodrigues, R., and J.~J. d.~A. Rangel. 2016.
\newblock ``Analysis of Ship Arrival Functions in Discrete Event Simulation
  Models of an Iron Ore Export Terminal''.
\newblock {\em Pesquisa
  Operacional\/}~36(1):45--66~\url{https://doi.org/10.1590/0101-7438.2016.036.01.0045}.


\bibitem[\protect\citeauthoryear{Trade and Development}{Trade and
  Development}{2024}]{unctad2024maritime}
Trade, U., and Development. 2024.
\newblock ``Review of Maritime Transport 2024''.
\newblock Technical report, United Nations Conference on Trade and
  Development~\url{https://doi.org/10.18356/9789211065923}.
\newblock Chapter IV: Port Performance and Vessel Calls. Container vessel size
  distributions from AIS-based port call analysis.

\bibitem[\protect\citeauthoryear{van Asperen, Dekker, Polman, {de Swaan Arons},
  and Waltman}{van Asperen et~al.}{2003}]{vanAsperen2003arrival}
van Asperen, E., R.~Dekker, M.~Polman, H.~{de Swaan Arons}, and L.~Waltman.
  2003.
\newblock ``Arrival Processes for Vessels in a Port Simulation''.
\newblock Technical Report ERS-2003-067-LIS, ERIM Report Series, Erasmus
  University Rotterdam.

\bibitem[\protect\citeauthoryear{Yu, Tang, Song, Yu, Qi, Li, and Zhang}{Yu
  et~al.}{2018}]{yu2018ship}
Yu, J., G.~Tang, X.~Song, X.~Yu, Y.~Qi, D.~Li {\em et~al}. 2018.
\newblock ``Ship Arrival Prediction and its Value on Daily Container Terminal
  Operation''.
\newblock {\em Ocean
  Engineering\/}~157:73--86~\url{https://doi.org/10.1016/j.oceaneng.2018.03.038}.


\bibitem[\protect\citeauthoryear{Yun and Choi}{Yun and
  Choi}{1999}]{yun1999simulation}
Yun, W.~Y., and Y.~S. Choi. 1999.
\newblock ``A Simulation Model for Container-Terminal Operation Analysis Using
  an Object-Oriented Approach''.
\newblock {\em International Journal of Production
  Economics\/}~59(1--3):221--230~\url{https://doi.org/10.1016/S0925-5273(98)00213-8}.


\end{thebibliography}

\section*{AUTHOR BIOGRAPHIES}

\vspace{3.8pt}

\noindent\textbf{QIAOHONG LI} is a Senior Research Fellow in the Department of Industrial Systems Engineering and Management (ISEM) 
at the National University of Singapore (NUS).
Her research interests include maritime studies, machine learning, and operations research. 
Her email address is \href{mailto:qhli@nus.edu.sg}{qhli@nus.edu.sg}.

\bigskip

\noindent\textbf{HAOBIN LI} is a Senior Lecturer in the Department of Industrial Systems Engineering and Management (ISEM) 
at the National University of Singapore (NUS), and Co-Director of the Centre of Excellence in Modelling and Simulation 
for Next Generation Ports (C4NGP). 
His research interests include discrete-event simulation modelling, simulation-based optimisation, with applications in port operations, 
and logistics. 
Dr Li is currently a Council Member of the INFORMS Simulation Society (I-SIM), and an overseas member of the Technical Committee
on Port and Shipping Economic Systems Engineering under the Systems Engineering Society of China (SESC). 
His email address is \href{mailto:li_haobin@nus.edu.sg}{li\_haobin@nus.edu.sg}.

\bigskip

\noindent\textbf{TIANHAO CHEN} is an IT Architect in the Department of Industrial Systems Engineering and Management (ISEM) 
at the National University of Singapore (NUS).
His research interests include port operations and logistics.  
His email address is \href{mailto:chth@nus.edu.sg}{chth@nus.edu.sg}.

\bigskip

\noindent\textbf{EK PENG CHEW} is a Professor in the Department of Industrial Systems Engineering and Management (ISEM) 
at the National University of Singapore (NUS), and Director of the Centre of Excellence in Modelling and Simulation 
for Next Generation Ports (C4NGP).   
His research primarily focuses on operations research, logistics and supply chain management, 
maritime and port optimization, and simulation-optimisation.
His email address is \href{mailto:isecep@nus.edu.sg}{isecep@nus.edu.sg}.

\end{document}